\documentclass[journal]{IEEEtran}

\usepackage{amsmath}
\usepackage{amsfonts}
\usepackage{amssymb}
\usepackage{mathtools}  
\usepackage{amsthm}
\usepackage{verbatim}
\usepackage{fontenc} 
\usepackage{graphicx}
\usepackage{mathrsfs}
\usepackage[]{algorithm2e}
\usepackage{booktabs}
\usepackage{subfigure}
\usepackage{tabulary}
\usepackage{booktabs}
\usepackage[justification=centering]{caption}
\usepackage{fancyhdr}
\usepackage{etoolbox}
\usepackage{comment}
\newtheorem{theorem}{Theorem}
\newtheorem{lemma}{Lemma}

\usepackage{xcolor}
\usepackage{hyperref}
\usepackage{soul}
\usepackage{graphicx}
\usepackage[nocompress]{cite}

\title{Architecture-Algorithmic Trade-offs in Multi-path Channel Estimation for mmWAVE Systems}

\author{\IEEEauthorblockN{Lyutianyang Zhang, Sumit Roy,  \IEEEmembership{Fellow, IEEE}, Liu Cao}\\
\IEEEauthorblockA{{Department of Electrical and Computer Engineering, University of Washington, Seattle, WA, USA} \\
Emails: \{lyutiz, sroy, liucao\}@uw.edu}}
\begin{document}
 \pagenumbering{gobble}
\maketitle
\begin{abstract}
5G mmWave massive MIMO systems are likely to be deployed in dense urban scenarios, where increasing network capacity is the primary objective. A key component in mmWave transceiver design is channel estimation which is challenging due to the very large signal bandwidths (order of GHz) implying significant resolved spatial multipath, coupled with large \# of Tx/Rx antennas for large-scale MIMO. This results in significantly increased training overhead that in turn leads to unacceptably high computational complexity and power cost. Our work thus highlights the interplay of transceiver architecture and receiver signal processing algorithm choices that fundamentally address (mobile) handset power consumption, with minimal degradation in performance. We investigate trade-offs enabled by conjunction of hybrid beamforming mmWave receiver and channel estimation algorithms that exploit available sparsity in such wideband scenarios. A compressive sensing (CS) framework for sparse channel estimation -  Binary Iterative Hard Thresholding (BIHT) \cite{jacques2013robust} followed by linear reconstruction method with varying quantization (ADC) levels- is explored to compare the trade-offs between bit-depth and sampling rate for a given ADC power budget. Performance analysis of the BIHT+ linear reconstruction method is conducted via simulation studies for 5G specified multi-path channel models and compared to oracle-assisted bounds for validation.
\end{abstract}

\section{Introduction}

A fundamental design objective for 5G networks is a significant increase in network capacity for urban hotspots. mmWave heterogeneous deployments - for both access and backhaul - have been proposed for such scenarios, with the goal of achieving Gbps rates per user. For instance, one can sense the ambient dominant reflectors to facilitate deploying the Access Point (AP) with good channel quality in network to achieve high data rate \cite{wei2017facilitating}. As is by now well-established, such high per user throughput is achieved by a combination of ultra-large signal bandwidth (order of GHz, available in mmWave bands) and large-scale MIMO transmission (with typical configurations such as $256$ (base station) x $16$ (client) antennas).  The currently active IEEE 802.11ay mmWave standard \cite{ghasem,11ay} has proposed multiuser massive MIMO using upto $60$ GHz channel bandwidth to achieve aggregate $100\ Gbs$ Wi-Fi. 

\begin{figure}[t]
  \centering
  \includegraphics[width=90mm, height=45mm]{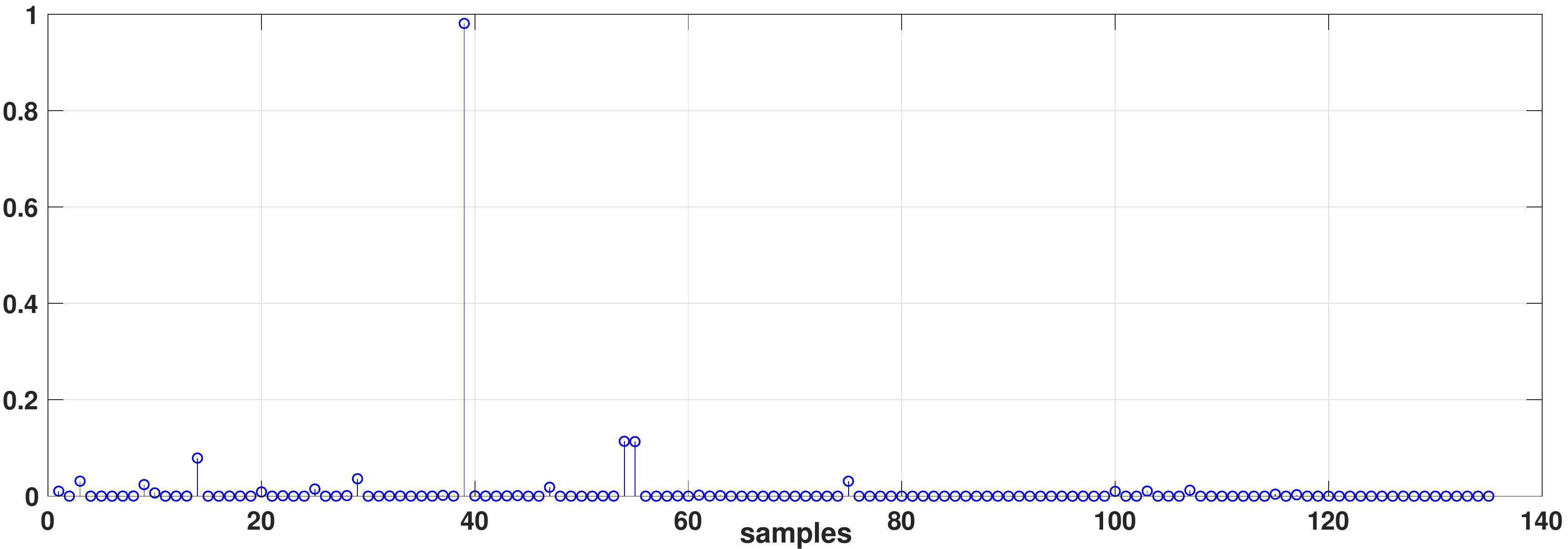}
\caption{Realization of IEEE 802.11ay multipath channel model \cite{channelmodel} in indoor environment; maximum delay spread is 55 ns @ tap spacing of 0.5 ns (sampling rate is 2GHz), resulting in $N=110$ channel length.}
\label{fig:intel}
\end{figure}

The ultra-wideband waveforms will result in multipath resolution at a fine scale \cite{raghavan2007capacity}. Hence, the time-domain channel impulse response has separable components at nano-second spacings, spread over the channel delay-spread duration. The recent measurement campaigns \cite{molisch2005ultrawideband,karedal2004statistical,chong2005modified} highlight a few key characteristics of mmWave multipath channels: i) these change on a much shorter timescale compared RF channels in response to client mobility and ii) the impulse response is significantly sparse (more so for indoor/residential environments). The mmWave outdoor channel has been studied extensively via measurements and consequent modeling \cite{rappaport2013millimeter}; the resulting average delay spread is found to be $T_{ds}=120$ ns. Thus for Nyquist sampling rate of $S=2$ Gsamples/s (for $1$ GHz channel), the discrete channel length is $\frac{T_{ds}}{S}=240$ taps. It has been shown that typical channels only have very few taps of significant value \cite{rappaport2013millimeter}, such as the realization of the 802.11ay multipath
channel shown in Fig.~\ref{fig:intel}, implying that the outdoor wireless channel is considerably sparse.   

%Mean (r.m.s) delay spread are 40.5468 (16.58) taps, for 0.5 ns tap spacing, at sampling rate of 2 GHz.
We next explore the two key dimensions underlying the analysis in this work. First, we consider  the power consumption budget for a mobile client, and review the compressed sensing algorithms as applied to  multi-path MIMO channel estimation.
\vspace{-2mm}
\subsection{Power constraints on ADC in mmWave MIMO systems}
\begin{figure*}[h]
  \centering
  \includegraphics[width=180mm, height=42mm]{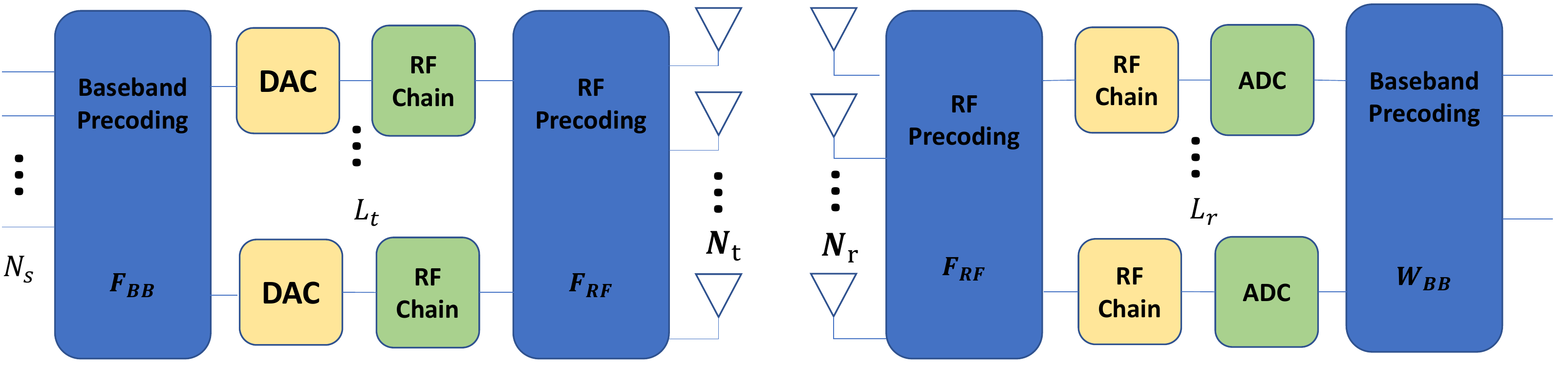}
\caption{Hybrid MIMO system architecture with ADC in each chain.}
\label{fig:intro}
\end{figure*}
\vspace{-1.5mm}
For MIMO at conventional cellular frequency bands (e.g. 2-3 GHz), transmit-side pre-coding is entirely realized in the digital domain to enable resulting interference cancellation between different data streams at a receiver. For such digital pre-coding, each antenna requires a dedicated energy-intensive radio frequency (RF) chain (including digital-to-analog converter and frequency up converter etc.), whose energy consumption is a significant part (about 250 mW per RF chain \cite{amadori2015low}) of the total energy consumption. If similar conventional digital precoding is implemented in mmWave massive MIMO system with a  very large number of antennas and corresponding RF chains, it will result in unsustainable energy consumption, e.g., 16 W for a massive MIMO system with 64 antennas. Clearly, this points to the need for fundamentally new approaches to downlink design whereby power consumption or UE lifetime is a main driver of the overall design approach. 

Following the above, we investigate trade-offs within a hybrid (digital-analog) system architecture \cite{gao2016energy} to fundamentally address the power consumption issue for large-scale mmWave MIMO systems. A hybrid MIMO system \cite{gao2016energy} - shown in Fig.\ref{fig:intro} -  associates one RF chain/ADC per {\em sub-array} (subset of antennas) - leading to $L_t (L_r)$ mixed signal chains at the transmit(receive) ends. This results in significant savings in power consumption and hardware cost. For example, while a traditional MIMO system with 16 RF chain for 16 antennas will incur $16 \times 250\ mW=4\ W $, $4 \times 4$ sub-arrays require only 4 RF chains leading to $4 \times 250\ mW=1\ W $.  Moreover, the bit-depth of the ADC as well as its sampling rate are  variables that can be further explored within such hybrid MIMO system to establish new performance-power operating points.  

For ADC power dissipation in terms of the sampling rate $f_s$ and bits/sample $B$, we adopt following \cite{choi2017adc}, 
\vspace{-0.1in}
\begin{equation} \label{Power_constriant}
P=c\,f_s 2^B 
\end{equation}
where $c=494$ fJ \cite{orhan2015low} denotes the Walden's constant, $f_s$ is the ADC sampling frequency, and $B$ is the bit-depth. This power constraint model yields equivalent energy $ E = P \,T = c\,M \, 2^B $ where $T$ is sampling duration, and $M = T \, f_s$ is the sample number. Our work will explore the choice of optimal sampling number $M$ and bit-depth $B$ given a fixed energy/power cost resulting from a prominent signal processing component - multi-path channel estimation.

\subsection{Compressive Channel Estimation}

Compressed sensing (CS) has emerged as a powerful tool to recover/estimate signals that satisfy a notion of sparsity\footnote{The degree of sparsity $K$ is defined as the number of non-zero entries in the signal vector $\beta \, \in \, {\cal R}^N$, i.e. the $l_0$ vector norm. All compressive sensing techniques rely on $ K \,  << \, N$}. As applied to the sparse channel estimation problem, well-known results in the literature state that as long as the signal design satisfies well-known restricted isometry properties (RIP), a sufficiently {\em small} number of measurements (determined by the signal dimension $N$ and the degree of sparsity $K$) suffices for arbitrarily accurate channel estimation \cite{eldar2012compressed}. CS algorithms has found applications in several communication sub-systems, including efficient image acquisition on a mobile phone via CS based decompression algorithms \cite{schneider2013new} and dynamic spectrum access via cognitive radios that require rapid (typically frequency domain) estimation of channel occupancy/availability \cite{tian2007compressed}. 

As already noted, the time-domain channel impulse response vector for mmWave multipath channels is significantly sparse and ripe for application of time-domain CS based estimation \cite{bajwa2010compressed,paredes2007ultra}\footnote{Note that the channel is {\em not} sparse in frequency domain; hence the traditional frequency-domain channel estimation approaches for OFDM or MIMO-OFDM modulation are unable to exploit sparsity.}. For 5G downlink, the signal recovery algorithm implemented on the UE must achieve both low latency and improved power efficiency. Among the total power budget for a mobile handset, the Analog-to-Digital converter is the most expensive component \cite{poulton200320}. Hence, \cite{panzner2014deployment} studied how to improve the overall transceiver power efficiency via the ADC bit-depth vs sampling frequency trade-offs. The other contributor to the power budget is the algorithmic complexity for channel estimation - the accuracy of channel estimation has known downstream impact on the ultimate decoder output. Hence accurate Channel State Information (CSI) potentially incurs large algorithmic complexity that in turn contributes to battery drain, and also be factored into the trade-off. 

{\em Our formulation is distinctive from prior art due to the focus on DL MU-MIMO systems within the hybrid analog-digital architecture, that allows exploration of ADC parameters within an energy budget perspective. The proposed systems design is backed by new results for bound on channel reconstruction SNR for an ideal (oracle-assisted) algorithm.} There exist prior approaches to {\em efficient (lower complexity)} MIMO channel estimation via exploitation of CS approaches but only a very limited number explore the impact of ADC quantization. CS is explored in \cite{alkhateeby2015compressed} for MU-MIMO downlink channel estimation without a quantizer. Continuous basis pursuit (CBP) was applied and two novel low-complexity algorithms proposed to exploit channel sparsity in \cite{sun2017millimeter}. Bussgang Linear Minimum Mean Squared Error (B-LMMSE) was applied to channel estimation for one-bit Massive MIMO systems in \cite{li2017channel}. Time-domain channel estimation with few bit ADC based Expectation Maximization plus generalized AMP method was proposed in \cite{mo2018channel}. Our work is thus distinctive from all the above in using a different CS based channel estimation approach in conjunction with choice of ADC parameters (bit-depth and sampling rate). 

 Since efficient CS channel estimation is enabled by choice of pilot sequences, we explore this aspect to discover additional algorithmic efficiencies. For example, \cite{gao2016channel} proposed a structured compressed sensing (SCS)-based channel estimation using non-orthogonal pilot design for downlink channel estimation that exploits {\em angular sparsity} of mmWave channel. Similarly, \cite{haupt2010toeplitz} proposed design of pilot sequences for SISO channel estimation, such that the resulting channel estimate is robust to additive Gaussian noise. However, these design did not exploit the Toeplitz nature of the observation matrix\footnote{In traditional CS, the measurement matrix is typically randomly chosen - e.g. i.i.d Gaussian entries - to satisfy the Random Isometry Property (RIP) condition.} that can be further utilized to obtain a more efficient MU-MIMO channel estimator. In our work, the pilot design from \cite{haupt2010toeplitz} is modified to suit the additional structural information inherent in our formulation, leading to improved channel reconstruction. 

In summary, the fundamental contributions of this work are two-fold: 1)	A mathematical model for MU-MIMO system with ADC quantization is proposed and a Restricted Isometry Property (RIP) for stable channel estimation for this MU-MIMO system model is derived that exploits the Toeplitz observation matrix structure; in turn, this is used to obtain an error bound for oracle-assisted linearly reconstructed channel estimation; 2) Explore choice of optimal ADC parameters (sampling number and bit-depth) given a fixed energy budget, thereby demonstrating feasible design solutions for hybrid mmWave transceiver architecture used in large-scale MIMO. 

The paper is organized as follows: based on the SISO system model \cite{haupt2010toeplitz}, a new model is proposed for MIMO multipath channel estimation with quantization effects. The pilot design \cite{haupt2010toeplitz} is extended to accommodate the new MU-MIMO model in conjunction with a 2-stage approach based on BIHT\cite{jacques2013robust} followed by an linear reconstruction estimation algorithm. The algorithm performance is then evaluated by simulations of the SISO and hybrid MIMO systems for various bit-depth and sampling number combinations to identify `optimal' transceiver designs. All key notations are summarized in Table I.

\begin{table}[t]\caption{Notations}
\begin{center}% used the environment to augment the vertical space
% between the caption and the table
\begin{tabular}{r c p{6.5cm} }
\toprule
$B$ & $=:$  & bit-depth of ADC\\
$M$ & $=:$ &  ADC sample number \\
$N$ & $=:$ &  channel vector dimension\\
$N_t$ & $=:$ & \# of transmitting antenna\\
$N_r$ & $=:$ & \# of receiving antenna\\
$K$ & $=:$ & \# of non-zero entries (degree of sparsity) of channel vector\\
$P$ & $=:$ & $M+N-1$ which is the length of training sequence \\
$\mathbf{X}$ & $=:$ & pilot/training matrix in SISO\\
$\mathcal{X}$ & $=:$ & pilot/training matrix in MIMO\\
$\mathbf{h}$ & $=:$ & channel vector in SISO\\
$\mathbf{\tilde{h}}$ & $=:$ & channel vector in MIMO\\
$\mathbf{y}$ & $=:$ & received signal before quantization in SISO\\
$\mathbf{y}_Q$ & $=:$ & received signal after quantization in SISO\\
$\mathbf{\tilde{y}}$ & $=:$ & received signal before quantization in MIMO\\
$\mathbf{\tilde{y}}_Q$ & $=:$ & received signal after quantization in MIMO\\
$\mathbf{X}^T$ & $=:$ & transpose of matrix $\mathbf{X}$\\
$\delta_K$ & $=:$ & RIP constant for $K$ sparse signal\\
$\lambda_{min}(\bullet)$ & $=:$ & minimum eigenvalue\\
$\mathbf{h}_{ij}$ & $=:$ & channel vector $h$ between transmitter $i$ - receiver $j$ pair\\
$\Sigma$ & $=:$ & $E[\mathcal{X}|_{\Omega}^{\dagger}\mathbf{\tilde{e}}]$\\
$\mathbf{X}_{R}$ & $=:$ &
real part of $\mathbf{X}$\\
$\mathbf{X}_{I}$ & $=:$ &
imaginary part of $\mathbf{X}$\\
\hline
\end{tabular}
\end{center}
\label{tab:TableOfNotation}
\end{table}

\section{Review: CS Based MIMO Channel Estimation}

Following prior formulations, the received signal ${\bf y} \in \mathcal{R}^{N_rM \times 1}$ for the hybrid MIMO system can be written in the following vector-matrix form  
\begin{equation}
\mathbf{\tilde{y}}=\mathcal{X}\mathbf{\tilde{h}}+\mathbf{\tilde{e}} ,
\end{equation}
where $\mathbf{\tilde{h}} \in \mathcal{R}^{N_tN_rN}$ is the channel vector to be estimated, and $\mathcal{X} \in \mathcal{R}^{N_rM \times N_tN_rN}$ composed of elements of the training sequence, is appropriately designed. The channel vector $\mathbf{\tilde{h}}$ is assumed $N_tN_rK$-sparse, i.e. $||\mathbf{h} ||_{0}:=supp(\mathbf{h}) \leq {N_tN_rK}$.   
The channel estimation problem can be expressed as 
\begin{equation}
\mathbf{\tilde{h}}_{est}=\min\limits_{\mathbf{\tilde{h}} \in \mathcal{R}^{N_tN_rN}}||\mathbf{\tilde{h}}||_{1}~~~s.t. ~||\mathbf{\tilde{y}}-\mathcal{X}\mathbf{\tilde{h}} ||<\epsilon ,
\end{equation}
i.e., we seek to minimize the ${\cal l}_1$ norm of the estimate, for a bounded reconstruction error $||\mathbf{\tilde{y}}-\mathcal{X}\mathbf{\tilde{h}}||  < \epsilon$. 

We explore an efficient sparse-channel estimation approach via the two-step algorithmic approach outlined in \cite{laska2012regime}. The 1st step identifies the support of the sparse vector $\mathbf{h}$ - in our case by the BIHT \cite{jacques2013robust} algorithm; thereafter in the 2nd-step, only the columns of $\mathcal{X}$  corresponding  to non-zero entry locations in $\mathbf{h}$ are used to generate the sparse estimate via the following \textbf{linear reconstruction method}
\begin{equation}\label{eq:linear_reconstruction}
\mathbf{\tilde{h}}_{est}|_{\Omega}=\mathcal{X}|_{\Omega}^{\dagger} \mathbf{\tilde{y}}, ~~~\mathbf{\tilde{h}}_{est}|_{\Omega^{C}}=0,
\end{equation}
where $\Omega$ is the index-set of non-zero entries\footnote{$\Omega^{C}$ is the complement set to $\Omega$.} in $\mathbf{\tilde{h}}$, $\mathcal{X}|_{\Omega}$ denotes the submatrix formed by  the index of $\Omega$, and $\dagger$ is the Moore-Penrose pseudo-inverse.
To guarantee the robustness of the signal recovery via CS, the matrix $\mathcal{X}$ must satisfy a restricted isometry property (RIP) \cite{candes2006stable} with order $N_t N_rK$ and constant $\delta$, that represents a sufficient condition for robust sparse signal recovery, i.e.,
\begin{equation}\label{Eq:RIP}
(1-\delta_{N_tN_rK})||\mathbf{\tilde{h}} ||_2^2 \leq ||\mathcal{X}\mathbf{\tilde{h}} ||_2^2 \leq (1+\delta_{N_tN_rK})||\mathbf{\tilde{h}}||_2^2.
\end{equation}
If $\mathcal{X}$ meets the condition above for all $N_tN_rK$-sparse $\mathbf{\tilde{h}}$, then it implies that the received signal $\mathbf{\tilde{y}}$ will only deviate from $\mathbf{\tilde{h}}$ in norm by at most the RIP constant $\delta_{N_tN_rK}$. Eq. \eqref{Eq:RIP} can also be understood as implying that any $N_tN_rK$-column of $\mathcal{X}$ has all its singular values close to 1, i.e., the isometry property can be stated as
\begin{equation}
\begin{aligned}
1-\delta_{N_tN_rK} &\leq \lambda_{min}(\mathcal{X}|_{\Omega_{N_tN_rK}}^{H}\mathcal{X}|_{\Omega_{N_tN_rK}})\\&\leq \lambda_{max}(\mathcal{X}|_{\Omega_{N_tN_rK}}^{H}\mathcal{X}|_{\Omega_{N_tN_rK}})\\&\leq 1+\delta_{N_tN_rK},
\end{aligned}
\end{equation}
where $\mathcal{X}|_{\Omega_{N_tN_rK}}$ denotes any submatrix with $N_tN_rK$ columns of original $\mathcal{X}$ and $|\Omega_{N_tN_rK}|=N_tN_rK$. RIP ensures stable linear reconstruction when the additive error $\mathbf{\tilde{e}}$ is white with variance  $\sigma_e^2$ - for which it is known that the mean-squared estimation error for an linear reconstruction estimator of the channel vector $\mathbf{\tilde{h}}$, is bounded as per 
\cite{davenport2012pros}, 
\begin{equation}
\begin{aligned}
\frac{N_tN_rK\sigma_e^2}{1+\delta_{N_tN_rK}} \leq E(|| \mathbf{\tilde{h}}-\mathbf{\tilde{h}}_{est} ||_2^2) \leq \frac{N_tN_rK \sigma_{e}^2}{1-\delta_{N_tN_rK}}.
\end{aligned}
\end{equation}

\section{MIMO System Model with Quantization}
\begin{figure}[t]
  \centering
  \includegraphics[width=90mm, height=50mm]{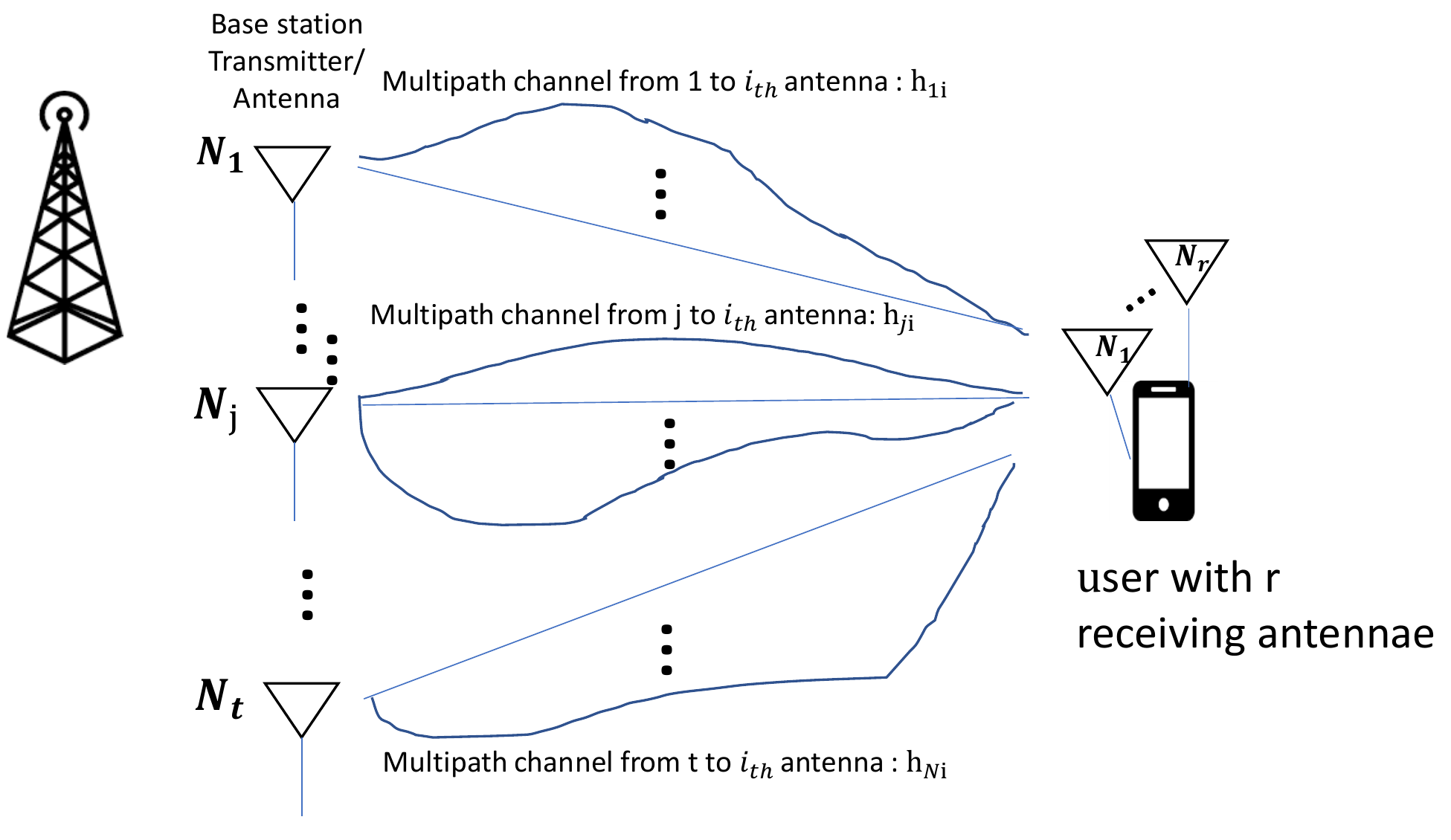}
\caption{Multipath channel in MIMO: each pairwise channel $\mathbf{h}_{ij}$, with $i=1,\dots,N_t$, $j=1,\dots, N_r$ is comprised of multipath elements, representing a time delay and phase shift.}
\label{fig:MIMO}
\end{figure}
 Fig.\ref{fig:MIMO} shows the multipath channel between a pair of transmitting and a receiving antenna. In this section, we first develop the system model for the SISO case, which is then stacked to formulate the MIMO model for a hybrid system with (few bit) ADC per chain. The base-band and RF precoding matrices of transmitting antenna and combining matrices at the receiving antenna are assumed to be identity matrices (corresponding to omni beam patterns). It is noteworthy that this assumption is actually a common approach for initial channel estimation stage in cross-layer design, for example, precoding matrix is set as an identity matrix for low power consumption in 5G NR V2X sidelink, according to 3GPP standard \cite{3GPP_TS_38211}.
\vspace{-2.07mm}
\subsection{SISO System Model}

Let $\mathbf{x}=\{x_i\}_{i=1}^P$ denote the training sequence of length $P$ transmitted from one antenna, as the input to a wireless channel characterized by a finite impulse response vector $\mathbf{h} \in \mathbb{C}^N$. The resulting observations $y \in  \mathbb{C}^{N+P-1}$ is the discrete-time convolution between the training/pilot signal $\mathbf{x}$ and the channel impulse response $\mathbf{h}$, in the presence of additive noise $\mathbf{e}$; i.e., $\mathbf{y}=\mathbf{x} \ast \mathbf{h}+\mathbf{e}$, where $ \ast$ denotes convolution and $\mathbf{e} \sim \mathcal{CN}(0,\sigma^2\mathbf{I}_{M})$. For general channel estimation without any sparsity, it is necessary that $p \geq n$. Stacking the observations for $M$ measurement instances yields the following input-output relation: 
\begin{equation} \label{eq:SISO}
\begin{aligned}
\begin{bmatrix}
y_1\\
\vdots\\
y_{M}
\end{bmatrix}=&
\begin{bmatrix}
x_N &	x_{N-1} & \dots  & x_1\\
x_{N+1} &	x_{N} & \dots & x_2\\
\vdots  &\vdots &\vdots &\vdots \\
x_{N+M-1} &	x_{N+M-2} & \dots & x_{M}\\
\end{bmatrix}
\begin{bmatrix}
h_1\\
\vdots\\
h_{N}
\end{bmatrix}+
\begin{bmatrix}
e_1\\
\vdots\\
e_M
\end{bmatrix}.
\end{aligned}
\end{equation}
that can be compactly written in vector/matrix form as
\begin{equation}\label{eq:siso}
\mathbf{y}=\mathbf{X}\mathbf{h}+\mathbf{e}.
\end{equation}
After the quantizer, the Eq. \eqref{eq:siso} is expressed as
\begin{equation} \label{eq: sisomodel}
\mathbf{y}_{Q}=Q_{B}\{\mathbf{X}\mathbf{h}+\mathbf{e}\},
\end{equation}
where $Q_{B}$ denotes the quantization function with bit-depth $B$, $\mathbf{y}_Q \in \mathcal{C}^{M}$, $\mathbf{X} \in \mathcal{C}^{M \times N}$ denotes a Toeplitz matrix containing the training sequence, chosen to satisfy RIP$(K,\delta_K)$, $\mathbf{h} \in \mathcal{C}^{N}$ is a $K$-sparse channel vector, and $\mathbf{e} \in \mathcal{C}^{M} \sim \mathcal{N}(0,\sigma_{e}^2)$.  

\subsection{MIMO System Model}

 Hence based on SISO model, we stack Eq.\eqref{eq:siso} to arrive at a model corresponding to the MIMO scenario shown in Fig.\ref{fig:MIMO}. For MIMO system with $N_t$ transmitters and $N_r$ receivers, this yields the system model
\begin{equation} \label{eq:MIMO}
\begin{aligned}
\mathbf{\tilde{y}}=\mathcal{X}\mathbf{\tilde{h}}+\mathbf{\tilde{e}},
\end{aligned}
\end{equation}
where
\begin{equation}
\begin{aligned}
&\mathcal{X}=\begin{bmatrix}
\mathbf{X}_1&\dots&\mathbf{X}_{N_t}&\\
\vdots&\vdots&\vdots&\mathbf{X}_1&\dots&\mathbf{X}_{N_t}\\
&&&\vdots&\vdots&\vdots\\
&&&&&&\mathbf{X}_1&\dots&\mathbf{X}_{N_t}\\
\end{bmatrix} \\ \\
&\normalsize
\mathbf{\tilde{h}}=[
\mathbf{h}_{11}
,\dots,
\mathbf{h}_{N_{t1}}
,\dots,
\mathbf{h}_{1N_r}
,\dots,
\mathbf{h}_{N_tN_r}
]^{T}_{N_tN_rN \times 1} \\ \\ &\mathbf{\tilde{y}}=[ \mathbf{y}_{1}, \dots, \mathbf{y}_{N_r}]^{T}_{N_rM \times 1}
\; \; \;\; \; \mathbf{\tilde{e}}=[\mathbf{e}_1, \dots, \mathbf{e}_{N_r}]^{T}_{N_rM \times 1}
\end{aligned}
\end{equation} 
\begin{comment}
\\&=
[
y_{11}
,\dots,
y_{1M}
,\dots,
y_{N_r1}
,\dots,
y_{N_rM}
]^{T}_{N_rM \times 1},\\
\end{comment}
\normalsize
where $\mathcal{X} \in \mathcal{C}^{N_rM \times N_tN_rN}$ is a block diagonal matrix, comprised of each submatrix $\mathbf{X}_i$ constructed from the training sequence sent from $i^{th}$ transmitting antenna. It is clear that the MIMO system model is a concatenated version of SISO model where each $\mathbf{h}_{ij} \in \mathcal{C}^{N}$ denotes a time domain channel vector between transmitter $i$ and receiver $j$ pair, that are individually $K$-sparse, and $y_{i} \in \mathcal{C}^{1}$ denotes the received signal at $t^{th}$ receiving antenna.  At the receiving antenna post ADC, the quantized system model can be expressed as 
\begin{equation}\label{eq:qMIMO}
\mathbf{\tilde{y}}_Q=Q_B\{\mathbf{\tilde{y}}\}=Q_B\{\mathcal{X}\mathbf{\tilde{h}}+\mathbf{\tilde{e}}\}.
\end{equation}

We now highlight how our model formulation deviates from prior art. First, it is noteworthy that the channel vector $\mathbf{\tilde{h}}$ has a special sparsity structure since now {\em each component} $\mathbf{h}_{ij}$ is $K$-sparse. Further, our formulation is in terms of complex quantities whereas the intended algorithm that we wish to apply - BIHT \cite{jacques2013robust} - has been developed for only real-valued matrices/vectors. Thus we need to re-cast our problem (by extracting the imaginary part and the real parts of the received signal) into an equivalent version that can bootstrap on prior results. \\

Denote $\mathbf{\tilde{y}}_1=\mathcal{X}\mathbf{\tilde{h}}_{R}+\mathbf{\tilde{e}}_{R}$, $\mathbf{\tilde{y}}_2=\mathcal{X}\mathbf{\tilde{h}}_{I}+\mathbf{\tilde{e}}_{I}$. Then it is obvious that $\mathbf{\tilde{y}}=\mathbf{\tilde{y}}_1+\mathit{i}~ \mathbf{\tilde{y}}_2$ from which we
have the following based on Eq. \eqref{eq:qMIMO}, 
\begin{equation*}\label{RIPofcomplexsiso}
\begin{aligned}
&Q_B\Bigg\{\begin{bmatrix}
\mathbf{\tilde{y}}_{R}\\\mathbf{\tilde{y}}_{I}
\end{bmatrix}\Bigg\}
\\= \ &Q_B\Bigg\{\begin{bmatrix}
real(\mathbf{\tilde{y}}_1)-img(\mathbf{\tilde{y}}_2)\\
img(\mathbf{\tilde{y}}_1)+real(\mathbf{\tilde{y}}_2)\\
\end{bmatrix}\Bigg\}
\\= \ &Q_B\Bigg\{\begin{bmatrix}
\mathcal{X}_{R}&-\mathcal{X}_{I}\\
\mathcal{X}_{I} &\mathcal{X}_{R}\\
\end{bmatrix}
\begin{bmatrix}
\mathbf{\tilde{h}}_{R}\\
\mathbf{\tilde{h}}_{I}\\
\end{bmatrix}+\begin{bmatrix}
\mathbf{\tilde{e}}_{R}\\
\mathbf{\tilde{e}}_{I}\\
\end{bmatrix}\Bigg\},
\end{aligned}
\end{equation*}

This yields 
\begin{equation}\label{eq:ulti_form}
    \begin{aligned}
     &Q_B\{\mathbf{\tilde{y}}_{R}\}=Q_B\Bigg\{
\mathcal{X}_1\mathbf{h}^{'}+\mathbf{e}_1^{'}\Bigg\}\\
     &Q_B\{\mathbf{\tilde{y}}_{I}\}=Q_B\Bigg\{
\mathcal{X}_2\mathbf{h}^{'}+\mathbf{e}_2^{'}\Bigg\},\\
    \end{aligned}
\end{equation}
where $\mathcal{X}_1 = [\mathcal{X}_{R}~-\mathcal{X}_{I}]$, $\mathcal{X}_2 = [\mathcal{X}_{I}~\mathcal{X}_{R}]$, $\mathbf{e}_1^{'}=\mathbf{\tilde{e}}_{R}$, $\mathbf{e}_2^{'}=\mathbf{\tilde{e}}_{I}$, and $\mathbf{h}^{\prime}=[\mathbf{\tilde{h}}_{R} ~ \mathbf{\tilde{h}}_{I}]^{T}$.
Henceforth, the original complex channel estimation problem in Eq.~(\ref{eq:qMIMO}) is now reformulated in terms of the equivalent real-valued problem represented by Eq.~(\ref{eq:ulti_form}). 

Next, we show how - by utilizing a training sequence design from \cite{haupt2010toeplitz} and the knowledge of the sparsity structure of the vector $ \mathbf{\tilde{h}}$, any submatrix of $\mathcal{X}_1$ and $\mathcal{X}_2$ satisfies RIP condition and hence $\mathcal{X}_1$ and $\mathcal{X}_2$ also preserves the RIP condition.

\begin{theorem} \label{theorem1}
\cite{haupt2010toeplitz} For the SISO model in \eqref{eq: sisomodel}, if $\mathbf{X}^{\prime} \in \mathbb{R}^{ M \times N}$ has i.i.d. $\pm 1/ \sqrt{M}$ elements $x_i, i \in \{1,\dots,N+M-1\}$ with probability $1/2$, $\mathbf{X}^{\prime}$ satisfies $RIP(K,\delta_{K})$ condition with probability at least $1-\exp(-c_1 M/K^2)$, provided $M \geq c_2 K^2 \log(N)$ and $c_2 \geq \frac{96c^2}{\delta_{K}^2-32c_1c^2}$ for any $c_1 \leq \frac{\delta_K^2}{32c^2}$.
\end{theorem}

Note that $\mathcal{X}_{R}$ and $\mathcal{X}_{I}$ are structured matrices comprised of sub-matrices $\mathbf{X}^{'}_1 \; \ldots \; \mathbf{X}^{'}_{N_t} $ as follows
\begin{equation}
\begin{aligned}
\begin{bmatrix}
\mathbf{X}^{'}_1&\dots&\mathbf{X}^{'}_{N_t}&\\
\vdots&\vdots&\vdots&\mathbf{X}^{'}_1&\dots&\mathbf{X}^{'}_{N_t}\\
&&&\vdots&\vdots&\vdots\\
&&&&&&\mathbf{X}^{'}_1&\dots&\mathbf{X}^{'}_{N_t}\\
\end{bmatrix},
\end{aligned}    
\end{equation}

where $\mathbf{X}^{'}_{i} \in \mathcal{R}^{M \times N}, i=N_1, \dots, N_t$  satisfy the statistical $RIP(K,\delta_K)$ condition per Theorem 1. Hence the conditions for RIP for Eq. (13) above is given by the following lemma. 

\begin{lemma} \label{lemma_1}
(MISO: $N_r=1$) Suppose $\mathbf{X}^{'}_i$ satisfies $RIP(K,\delta_i)$ with $i=1,2,\dots, N_t$ and $\delta_K=\max\limits_i(\delta_i)$, then $\mathbb{X}=[\mathbf{X}^{'}_1,\mathbf{X}^{'}_2,\dots,\mathbf{X}^{'}_{N_t}]$, where $\mathbb{X} \in \mathcal{R}^{M\times N_tN}$, also satisfies $RIP (N_tK, \delta_K)$ for orthogonal training sequences. 

\begin{proof}
 Let $\Omega_i$ be an index set denoting a randomly chosen $K$ column subset for matrix $\mathbf{X}^{\prime}_i$, written as $\mathbf{X}^{\prime}_i|_{\Omega_i}$. The corresponding index set $\Omega$ representing $K\, N_t$ column subset of matrix $\mathbb{X}$ is then defined as follows:
\begin{equation}
\begin{aligned}
    \Omega&=[\Omega_1, \dots, \Omega_{N_t}]=\bigcup_{i=1}^{N_t}\Omega_i,\\
    \mathbb{X}_{\Omega}&=[\mathbf{X}^{'}_1|_{\Omega_1},\mathbf{X}^{'}_2|_{\Omega_2},\dots,\mathbf{X}^{'}_{N_t}|_{\Omega_{N_t}}]
    \end{aligned}
\end{equation}
%which satisfies $RIP (N_tK, \delta_K)$ corresponding to concatenated channel vector $\mathbf{\tilde{h}}$.

%Suppose $\mathbf{X}^{'}_i$ satisfies $RIP(K,\delta_i)$ with $i=1,2,\dots, N_t$, the sufficient and necessary conditions follows 
%$1-\delta_K \leq \lambda_{min}(\mathbf{X}^{'}_i|_{\Omega_i}^{H}\mathbf{X}^{'}_i|_{\Omega_i}) \leq \lambda_{max}(\mathbf{X}^{'}_i|_{\Omega_i}^{H}\mathbf{X}^{'}_i|_{\Omega_i}) \leq 1+\delta_K$ for any $|\Omega_i|=K$.
%remove the same idea applies -> from rip of each individuals
%define Sigma in table one and locally.

We assume that different training sequences $\mathbf{X}^{\prime}_i$ and $\mathbf{X}^{\prime}_j$ are independent for $i \neq j$. Since each element of $\mathbf{X}^{\prime}_i$ can only be $\pm\,1/\sqrt{M}$ with equal probability and two different training sequences are chosen orthogonal by design, i.e., $\mathbf{X}^{'}_i|_{\Omega_i}^{H}\mathbf{X}^{'}_j|_{\Omega_j} = \mathbf{0}$ when $i \neq j$ using the law of large number (L.L.N). Hence, ${\mathbb{X}_{\Omega}}^{H}\mathbb{X}_{\Omega}$ is (in the limit) a block diagonal matrix, i.e.,
\scriptsize
\begin{equation}\label{eq:lemma1_1}
{\mathbb{X}_{\Omega}}^{H}\mathbb{X}_{\Omega} \; \rightarrow \; 
\left(\begin{bmatrix}
\mathbf{X}^{'}_1|_{\Omega_1}^{H}\mathbf{X}^{'}_1|_{\Omega_1}\\
&\mathbf{X}^{'}_2|_{\Omega_2}^{H}\mathbf{X}^{'}_2|_{\Omega_2}\\
&&\vdots
&&\mathbf{X}^{'}_{N_t}|_{\Omega_{N_t}}^{H}\mathbf{X}^{'}_{N_t}|_{\Omega_{N_t}}
\end{bmatrix}\right) 
\end{equation}
\normalsize
and hence 
\scriptsize
\begin{equation}\label{eq:lemma1_1}
\begin{aligned}
&~\lambda_{min}({\mathbb{X}_{\Omega}}^{H}\mathbb{X}_{\Omega})
\\
&= \lambda_{min}\left(\begin{bmatrix}
\mathbf{X}^{'}_1|_{\Omega_1}^{H}\mathbf{X}^{'}_1|_{\Omega_1}\\
&\mathbf{X}^{'}_2|_{\Omega_2}^{H}\mathbf{X}^{'}_2|_{\Omega_2}\\
&&\vdots
&&\mathbf{X}^{'}_{N_t}|_{\Omega_{N_t}}^{H}\mathbf{X}^{'}_{N_t}|_{\Omega_{N_t}}
\end{bmatrix}\right)\\&=\min\limits_{i}\{\lambda_{min}( \mathbf{X}^{'}_i|_{\Omega_i}^{H} \mathbf{X}^{'}_i|_{\Omega_i}\},
\end{aligned}
\end{equation}
\normalsize
%for each $\mathbf{h}_{ij}$ is $K$-sparse and each $\Omega_i$ denotes the non-zero items for the corresponding N columns of $\mathbb{X}$. For example, a MIMO system with $N_t=2$ and $N_r=1$, $\mathbf{X}^{'}_1$ only has to guarantee the singular value to be close to 1 corresponding to the channel vector $\mathbf{h}_{11}$ without affecting $\mathbf{h}_{21}$. Hence, we have the following inequality

Now, since each $\mathbf{X}_i$ satisfies $RIP(K,\delta_i)$, we have the following \cite{candes2005decoding},
\begin{equation}
    \begin{aligned}
    1-\delta_K &\leq \min\limits_i(\lambda_{\min}(\mathbf{X}^{'}_i|_{\Omega_i}^{H}\mathbf{X}^{'}_i|_{\Omega_i}))\\& \leq \max\limits_i(\lambda_{\max}(\mathbf{X}^{'}_i|_{\Omega_i}^{H}\mathbf{X}^{'}_i|_{\Omega_i}))\\& \leq 1+\delta_K,
    \end{aligned}
\end{equation}
where $\min\limits_i(\lambda_{\min}(\mathbf{X}^{'}_i|_{\Omega_i}^{H}\mathbf{X}^{'}_i|_{\Omega_i})) =  \lambda_{min}({\mathbb{X}|_{\Omega}}^{H}{\mathbb{X}}_{\Omega})$ and $\max\limits_i(\lambda_{\max}({\mathbf{X}^{'}}_i|_{\Omega_i}^{H}\mathbf{X}^{'}_i|_{\Omega_i}))=\lambda_{max}({\mathbb{X}}|_{\Omega}^{H}\mathbb{X}|_{\Omega})$. Therefore, $\mathbb{X}$ satisfies $RIP(N_tK,\delta_K)$. 
\end{proof}
\end{lemma}

\begin{lemma}\label{lemma_2}
(MIMO) Suppose matrix $\mathbb{X}$ satisfies $RIP (N_tK, \delta_K)$ as in Lemma.\ref{lemma_1}, then block diagonal matrix
$\mathcal{X}=\begin{bmatrix}
\mathbb{X}\\
&\mathbb{X}\\
&&\ddots   \\
&&&\mathbb{X}\\
\end{bmatrix}$, where $\mathcal{X} \in \mathcal{R}^{N_rM \times N_rN_tN}$, satisfies $RIP(N_t N_r K, \delta_K)$.
\begin{proof}
% with the largest RIP constant $\delta_K$  out of $\mathbb{X}_i$, $i=1, \dots, N_r$. 
Same as in Lemma \ref{lemma_1}.
\end{proof}
\end{lemma}

Note that Lemma \ref{lemma_1} and Lemma \ref{lemma_2} are valid only under the condition that the channel vector $\mathbf{h}$ has the the special sparsity pattern -  each $\mathbf{h}_{ij}$ is itself a $K$-sparse vector. Combining Lemma \ref{lemma_1} and \ref{lemma_2}, we obtain the following theorem:
\begin{theorem}
If $\mathbf{X}^{'}_i$ satisfies $RIP(K,\delta_i)$ with $i=1,2,\dots, N_t$ and $\max\limits_i(\delta_{i})=\delta_K$, then $\mathcal{X}_1=[\mathcal{X}_{R}~~ -\mathcal{X}_{I}]$ and $\mathcal{X}_{2}=[\mathcal{X}_{I}~~ \mathcal{X}_{R}]$ both
satisfy $RIP(2N_tN_r K, \delta_K)$.

\begin{proof}
Based on Lemma \ref{lemma_1} and \ref{lemma_2}, $\mathcal{X}_{R}$ and $\mathcal{X}_{I}$ both satisfy RIP$(N_tN_rK,\delta_{K})$. Then we have the following
\begin{equation}\label{eq:theorem2_eigen}
\begin{aligned}
&~~~~\lambda_{min}(\mathcal{X}_{1}|_{\Omega}^{H}\mathcal{X}_{1}|_{\Omega})
\\&=\lambda_{min}([\mathcal{X}_{R}|_{\Omega_{R}}~ ~-\mathcal{X}_{I}|_{\Omega_{I}}]^{H}[\mathcal{X}_{R}|_{\Omega_R}~ ~-\mathcal{X}_{I}|_{\Omega_I}])
\\
&\stackrel{a}{=}\lambda_{min}\begin{bmatrix}\mathcal{X}_{R}|_{\Omega_R}^{H}\mathcal{X}_{R}|_{\Omega_R}& 0\\
0 &\mathcal{X}_{I}|_{\Omega_I}^{H}\mathcal{X}_{I}|_{\Omega_I}
\end{bmatrix}.
\end{aligned}
\end{equation}
The equality (a) holds because $\mathbf{X}_i$ and $\mathbf{X}_j$ are independent if $i\neq j$, which implies that $\mathcal{X}_R|_{\Omega_R}^{H}\mathcal{X}_I|_{\Omega_I}=\mathbf{0}$. It is obvious that the minimum eigenvalue in Eq.\eqref{eq:theorem2_eigen} is either $\lambda_{min}(\mathcal{X}_{R}|_{\Omega}^{H}\mathcal{X}_{R}|_{\Omega})$ or $\lambda_{min}(\mathcal{X}_{I}|_{\Omega}^{H}\mathcal{X}_{I}|_{\Omega})$. Meanwhile, the sparsity of $[\mathcal{X}_{R}|_{\Omega}~ ~-\mathcal{X}_{I}|_{\Omega}]$ is $2N_tN_rK$ and we denote the corresponding RIP constant as $\delta_{2N_tN_rK}$. Then, we have the following inequality:
\begin{equation}
\begin{aligned}
&1-\delta_{2N_tN_rK} \\&\leq 
\lambda_{min}([\mathcal{X}_{R}|_{\Omega_R} ~~\!-\mathcal{X}_{I}|_{\Omega_I}]^{H}[\mathcal{X}_{R}|_{\Omega_R}~~\!-\mathcal{X}_{I}|_{\Omega_I}])\\& \leq
\lambda_{max}([\mathcal{X}_{R}|_{\Omega_R}~~\!-\mathcal{X}_{I}|_{\Omega_I}]^{H}[\mathcal{X}_{R}|_{\Omega_R}~~\!-\mathcal{X}_{I}|_{\Omega_I}])\\ & \leq  1+ \delta_{2N_tN_rK}.
\end{aligned}
\end{equation}

Hence, given $\mathbf{\tilde{h}}_{R}$ and $\mathbf{\tilde{h}}_{I}$ are both $N_tN_rK$ sparse, matrix $\begin{bmatrix}
\mathcal{X}_{R}~ &-\mathcal{X}_{I} 
\end{bmatrix}$ has $RIP (2N_tN_rK,\delta_{2N_tN_rK})$ where $ \delta_{2N_tN_rK}$ is the maximum RIP constant between $\mathcal{X}_{R}$ and $\mathcal{X}_{I}$. Hence, $\delta_{2N_tN_rK}=\max\limits_{\{R,I\}}\delta_{N_tN_rK}$ with $\max\limits_{\{R,I\}}\delta_{N_tN_rK}$ denoting the maximum RIP constant between $\mathcal{X}_R$ and $\mathcal{X}_I$. Based on Lemma \ref{lemma_1} and \ref{lemma_2}, $\max\limits_{\{R,I\}}\delta_{N_tN_rK}=\delta_K$ is the largest RIP constant out of all submatrix $X^{\prime}_{i}$ for both $\mathcal{X}_{R}$ and $\mathcal{X}_{I}$. Hence, $\mathcal{X}_1$ and $\mathcal{X}_2$ both satisfy $RIP(2N_tN_rK,\delta_K)$.
\end{proof}

\end{theorem}

In summary, stable recovery for the MIMO system is guaranteed if we reformulate the reconstruction based on $\mathbf{\tilde{y}}_{R}$ and $\mathbf{\tilde{y}}_{I}$ and design real-valued training sequences for Eq. \eqref{eq:ulti_form}. It is now clear that $\mathcal{X}_1$ and $\mathcal{X}_2$ both satisfy $RIP(2N_tN_rK,\delta_K)$ and hence guarantees robust sparse channel recovery with noisy measurements.

\begin{theorem}\label{errorbound}
\cite{laska2012regime} Suppose that $Q_{B}(\mathcal{X}_{R}\mathbf{\tilde{h}}_{R}+\mathbf{\tilde{e}}_{R})$ and each element of the matrix are all real. Let the channel vector $\mathbf{\tilde{h}}_{R} \in \mathcal{R}^{N_tN_rN}$ be $N_tN_rK$-sparse. Let the non-zero elements in $\mathbf{\tilde{h}}_{R} \sim \mathcal{N}(0,\sigma_{\mathbf{h}}^{2})$, each element of the noise $\mathbf{\tilde{e}}_{R} \in \mathbb{R}^{N_rM}$ be a random white, zero-mean Gaussian vairable with variance $\sigma_{e}^{2}$, $\mathcal{X}_{R} \in \mathcal{R}^{N_rM \times N_tN_rN}$ satisfies $RIP(N_tN_rK,\delta_{N_tN_rK})$. Choose $Q_B$ to be the optimal scalar quantizer with $B>1$ that minimize the MSE for the distribution of the measurements $\mathcal{X}_{R}\mathbf{\tilde{h}}_{R}+\mathbf{\tilde{e}}_{R}$. Then, the reconstruction-SNR of the oracle-assisted \footnote{Oracle-assisted reconstruction is the linear reconstruction method in Eq.\eqref{eq:linear_reconstruction} with the assumption that the support vector location is perfectly known.} reconstruction for channel vector $\mathbf{\tilde{h}}_{est}$ satisfies
\begin{equation}\label{eq:errorbound}
\frac{||\mathbf{\tilde{h}}_R||^2_2}{E[||\mathbf{\tilde{h}}_{R}-\mathbf{\tilde{h}}_{est}||_2^2]} \leq
\frac{||\mathbf{\tilde{h}}_R||^2_2}{\frac{N_tN_rK}{1+\delta_{N_tN_rK}}\lambda_{min}(\Sigma)},
\end{equation}
where $\Sigma=E[\mathcal{X}_{R}|_{\Omega}^{\dagger}\mathbf{\tilde{e}}_{R}]$
\end{theorem}
\begin{proof}
Refer to \cite{laska2012regime} Lemma 1.
\end{proof}

In summary, we modified a key theorem from \cite{laska2012regime} to incorporate the Toeplitz property of the observation matrix, using results from \cite{haupt2010toeplitz} to guarantee RIP.  Note that the RIP result is only valid for stacked MIMO model in Eq. \eqref{eq:MIMO}, where the resultant channel vector is a concatenation of sparse sub-vectors. Theorem \ref{errorbound} is computed numerically in the next section to demonstrate the lower bound of the channel recovery error. 

%\begin{remark}
%Two training sequence design methods shown above are real training sequence value and imaginary training sequence value respectively. Hereby we show the superiority of real-valued training sequence over imaginary-valued training sequence. First of all, if we want to utilize imaginary-valued sequences, one must guarantee the  independence for the real part and imaginary for each elements, which is difficult to realize. However, for real-valued sequences, one only have to devise an i.i.d. sequence. It is also obvious that real-values sequences could reserve power compared to imaginary-valued sequences during modulation. Last but not least, the RIP constant in imaginary-valued scenario is larger than the real-valued, which means that less robust recovery for imaginary-valued training sequences. Hence, real-valued sequences are recommended to implemented because its performance is robust, easy to implement, and power-saving advantage. 
%\end{remark}

\begin{table*}[h]
\caption{ADC parameters for $20mW$ Power Constraint} 
\label{tab:SISO_parameter}
\begin{center}
{\renewcommand{\arraystretch}{1.1}
\begin{tabular}{c  c  c  c  c  c  c  c }
\hline
Bit-depth & 2 & 3 & 4 & 5 & 6 &7 &8\\
\hline 
Sampling frequency (GHz) $f_s$ & 10 & 5 & 2.5 & 1.25& 0.63  &0.31 &0.16\\ 
\hline 
Sampling number $M$ & 1000 & 500 & 250 & 125 & 75 &38 &19 \\ 
\hline 
\end{tabular}
} 
\end{center}
\end{table*}

\section{SISO \& MIMO Channel Estimation: Performance with BIHT + Linear Recontruction}

\begin{figure}[t]
\begin{center}
  %\subfigure[\small{BIHT Support Vector Recovery Method for Different Input SNR}]
 %{\resizebox{5.5cm}{!}{\includegraphics{fig/SISO_BIHT.eps}}}

\subfigure[\small{Optimum ADC Bit Size for Different Input SNR}]
{\resizebox{5.5cm}{!}{\includegraphics{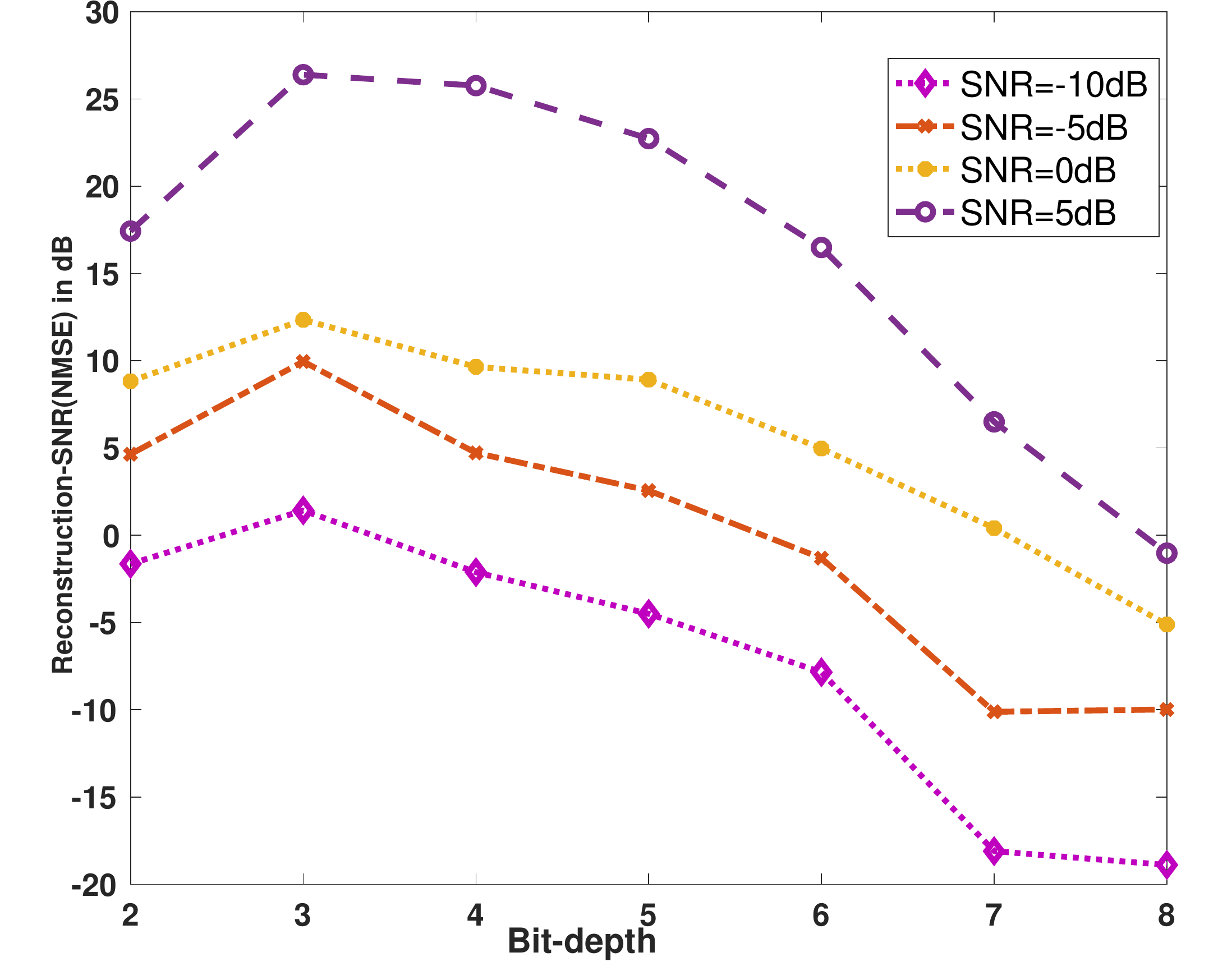}}}
   \subfigure[\small{Channel Estimation Performance for Different Input SNR }]
 {\resizebox{5.5cm}{!}{\includegraphics{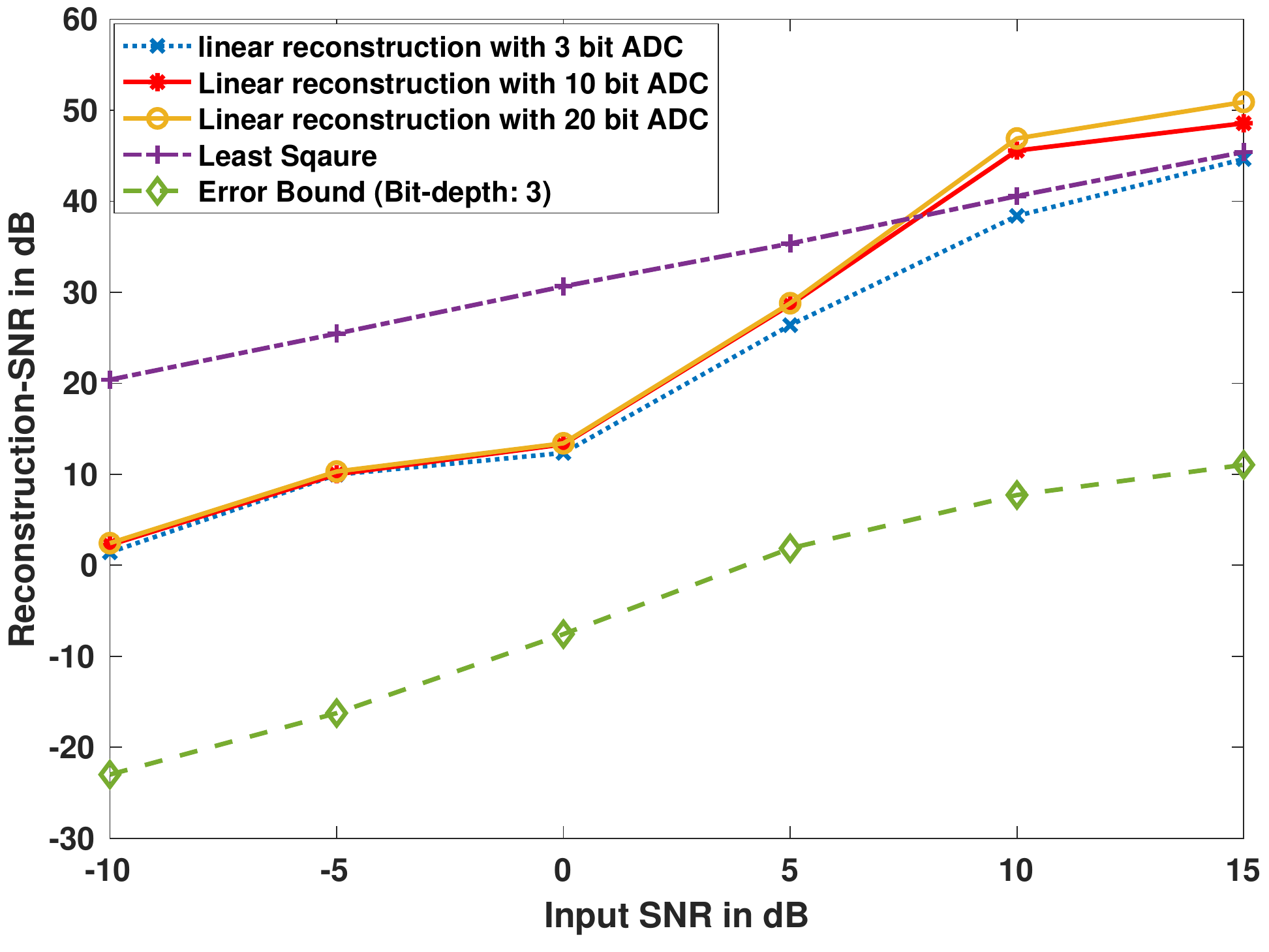}}}
\end{center}
   \caption{SISO multipath channel estimation using BIHT + Linear Reconstruction. Error bound is obtained using Eq.\eqref{eq:errorbound}}
\label{SISO_sim}
\end{figure}

We present simulation results for SISO and MIMO channel estimation scenarios following the system diagram Fig. \ref{fig:csdiagram}, to explore the optimal values for $B$ and $f_s$ under a given ADC power constraint, and subsequently compare  channel estimation performance (with the optimum choices of ADC parameters $B, f_s$) with respect to least squares and relevant methods such as EM-GM-VAMP\footnote{EM-GM-VAMP is an extended version of expectation-maximization method.}.

\vspace{-5.2mm}
\subsection{Simulation for SISO}

Binary Iterative Hard Thresholding (BIHT) (discussed in Appendix) is used to identify the non-zero locations of the sparse channel vector followed by linear reconstruction in Eq.\eqref{eq:linear_reconstruction} for final channel estimation, as shown in Fig. \ref{fig:csdiagram}. Note that the one-bit quantized received signal $sgn(\mathbf{\tilde{y}}_R)$ or $sgn(\mathbf{\tilde{y}}_I)$ suffices for BIHT whereas the subsequent linear reconstruction uses $B$-bit quantization. BIHT plus linear channel estimation is implemented to find the optimal ADC parameters given a fixed power constraint. According to \cite{orhan2015low}, high speed ADC power consumption lies in  $1- 50$ mW range. Hence we set ADC power budget to $20$ mW  for all simulations in this section for input SNRs of $-10, -5, 0, 10, 15$ dB. The expected values for the sparsity and channel maximum delay spread are $K=5$ and $200$, obtained from multiple realizations of IEEE 802.11ay multipath channel model\cite{channelmodel}. Hence,  channel vector $\mathbf{h}$ of length $N=200$ with $0.5$ ns tap spacing is chosen. The non-zero channel vector $\mathbf{h}$ components are distributed as $\mathcal{CN}(0,1)$,i.e., a standard complex normal distribution with the magnitude of the largest entry normalized to 1 to align with realizations of \cite{channelmodel} shown in Fig. \ref{fig:intel}.

Training sequence length is $500$ samples and based on Eq.\eqref{Power_constriant},  we have the sampling frequency vs Bit-depth in Table. \ref{tab:SISO_parameter}. Monte Carlo simulation with 1000 trials are carried out and the Reconstruction Signal-to-Noise Ratio (RSNR), defined below, is utilized as the performance metric: 
\begin{equation}
RSNR=\frac{||\mathbf{\tilde{h}} ||_2^{2}}{||\mathbf{\tilde{h}}-\mathbf{\tilde{h}}_{est}||_2^2},
\end{equation}
where $\mathbf{\tilde{h}}_{est}$($\mathbf{\tilde{h}}$) denotes the estimated (true) channel vector, respectively.

The simulation results for SISO are shown in Fig.\ref{SISO_sim}. As expected,  channel estimation performance decreases with SNR. For SNR below $5$ dB, optimal bit-depth and sampling frequency is $3$ and $5$ GHz, respectively; for SNR of $15$ dB, the optimal bit-depth and sampling frequency is $4$ and $2.5$ GHz. Hence, under a given power constraint of $20$ mW, the simulation results show the optimal bit-depth lies between $3-4$  in the SNR range from $-10$ to $15$ dB. In Fig. 4(b), the reconstruction SNR as a function of ADC bit-depth shows the existence of an optimal value for different input SNRs. Fig. 4 shows that the proposed linear reconstruction with 3 bit ADC yields acceptable channel RSNR even at input SNR of $-5$ dB. Fig. 4  also shows the relationship of the optimal error bound  in Theorem \ref{errorbound} with 3-bit ADC to the results with our proposed method (BIHT+ linear reconstruction) with 3-bit ADC. Our method with 3-bit ADC significantly outperforms traditional least squares and approaches the error bound around input SNR of $5$ dB. Note that the performance of our method (BIHT plus linear-reconstruction) improves on the 3-bit error bound, but 10 and 20-bit ADCs outperform the 3-bit bound for high SNR, but at a significant power cost. Hence, 3-bit ADC in conjunction with BIHT+linear reconstruction provides a very satisfactory design solution - with a small compromise in output RSNR, {\em significant} savings on ADC power consumption is achieved.

\vspace{-4.8mm}
\subsection{Simulation for MIMO}
\begin{figure}[h]
\begin{center}
\subfigure[\small{2 x 1 channel estimation }]
{\resizebox{6cm}{!}{\includegraphics{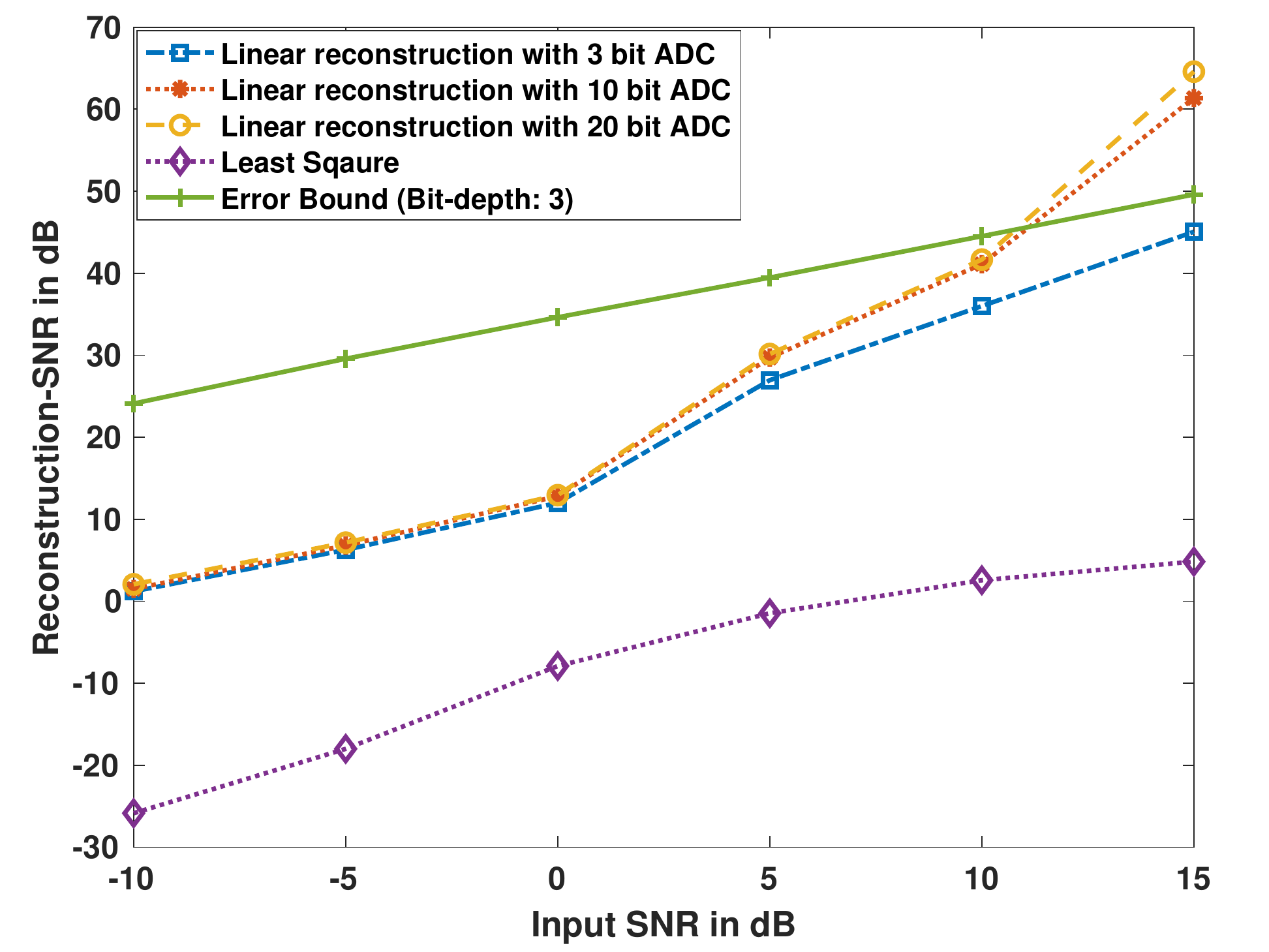}}}

\subfigure[\small{10 x 1 channel estimation }]
 {\resizebox{6cm}{!}{\includegraphics{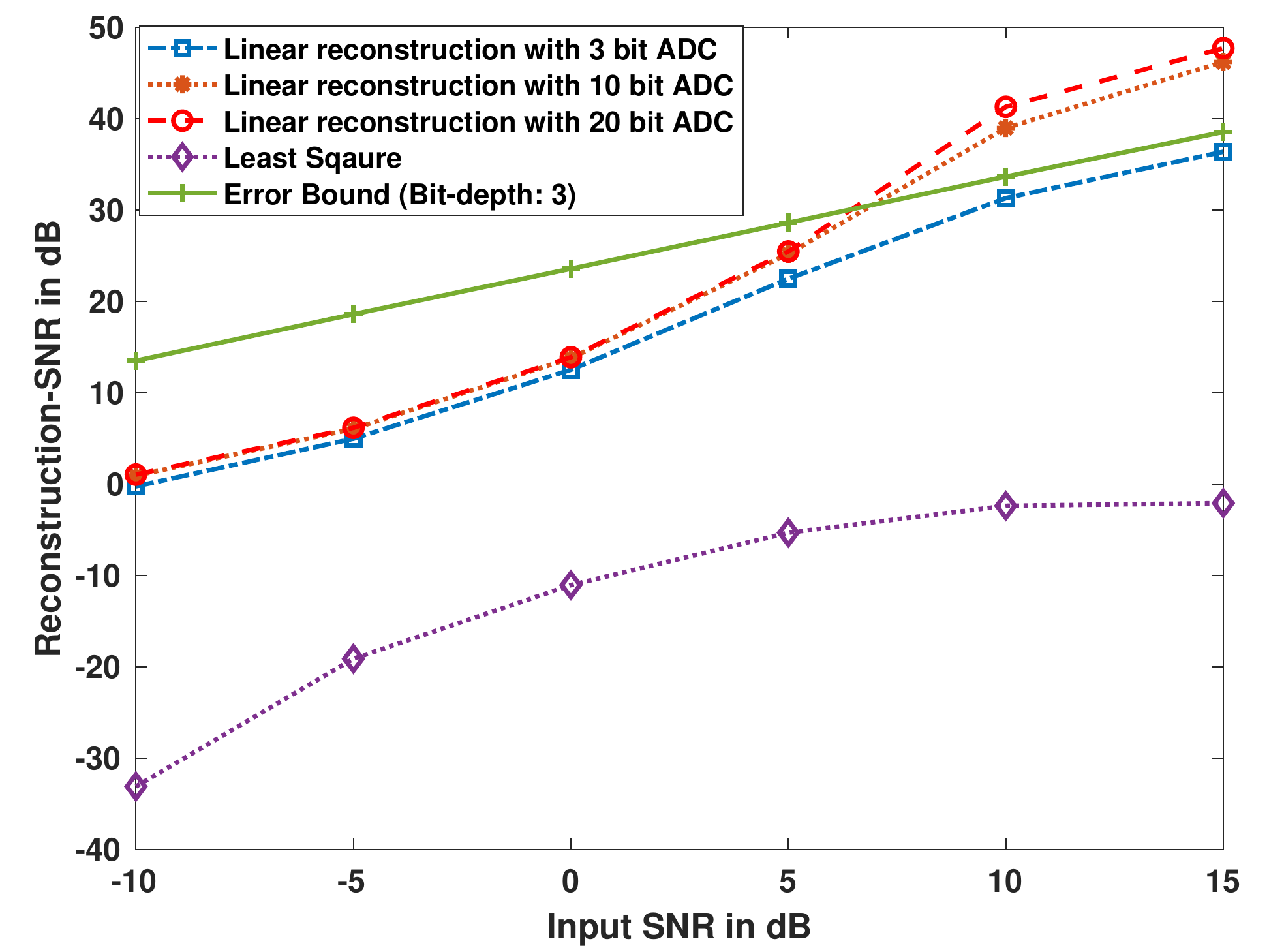}}}
  \subfigure[\small{4 x 2 channel estimation}]
 {\resizebox{6cm}{!}{\includegraphics{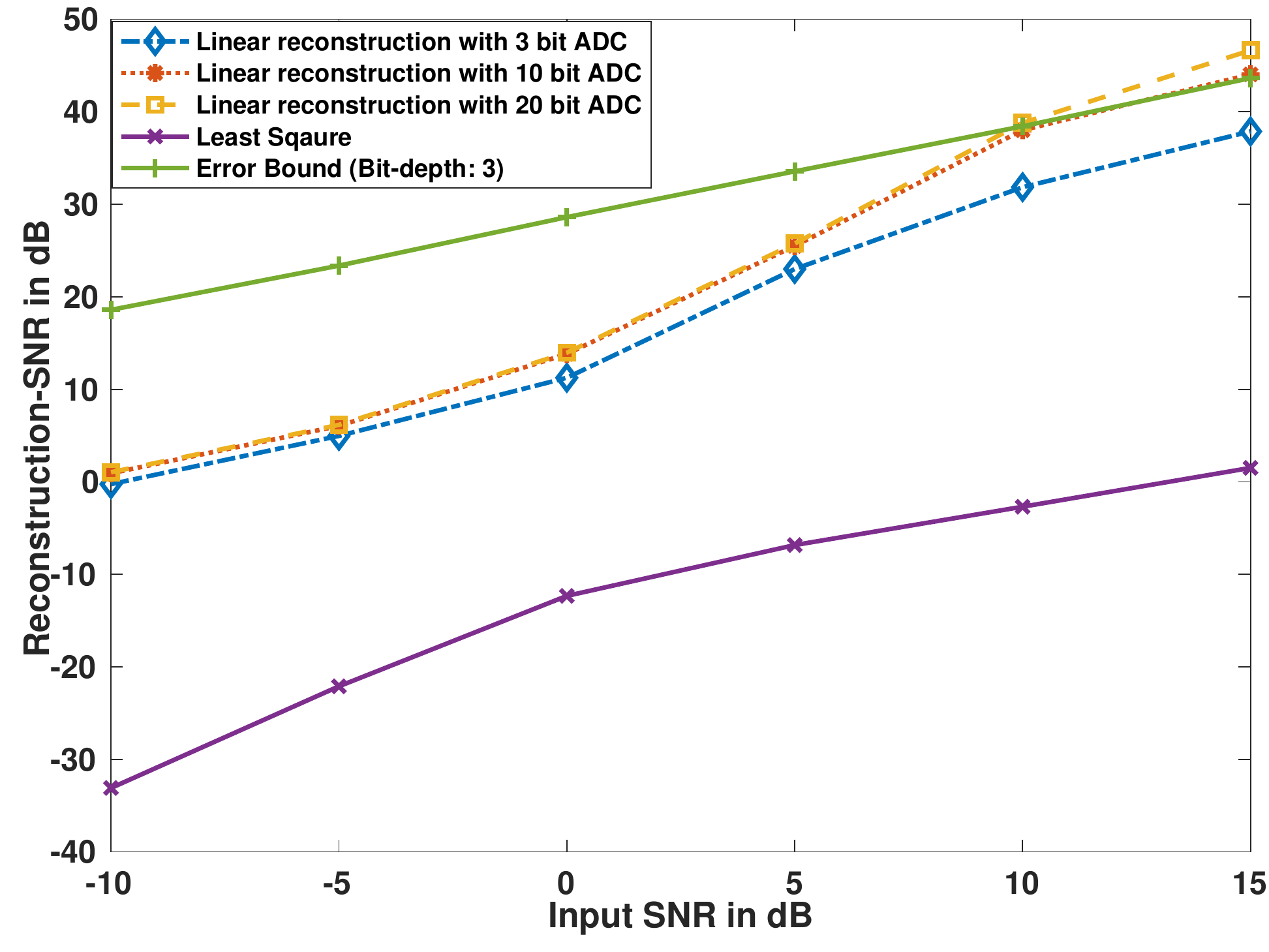}}}

\end{center}
   \caption{Simulation for MIMO multipath channel estimation using BIHT + Linear reconstruction method. Error bound is obtained using Eq.\eqref{eq:errorbound}}
\label{MIMO_sim}
\end{figure}

Next, simulations for MIMO systems are implemented under the same $20$ mW power constraint. Since the channel dimension is now larger, more data must be utilized, and hence the sampling duration increases proportionally. Namely, the sampling durations for $ 2 \times \, 1$,$ 4\,\times\, 2$ and  $ 10 \,\times 1$ MIMO system is $200, 800$ ns and  $1000$ ns, respectively. The training sequence length sent over each antenna is $500$ and hence $ 10\,\times 1$ MIMO system requires $5000$ training symbols.   The same methodology as SISO - BIHT followed by linear reconstruction - is used to identify the optimal bit-depth and the sampling number for ADC. The results are compared with the error bound for linear reconstruction; as we see from Fig.\ref{MIMO_sim}, the obtained RSNR performance is very close to the optimal with only a small gap  with 3 bit ADC, and is again much better than traditional least squares. 

\begin{figure}[h]
  \centering
  \includegraphics[width=70mm, height=50mm]{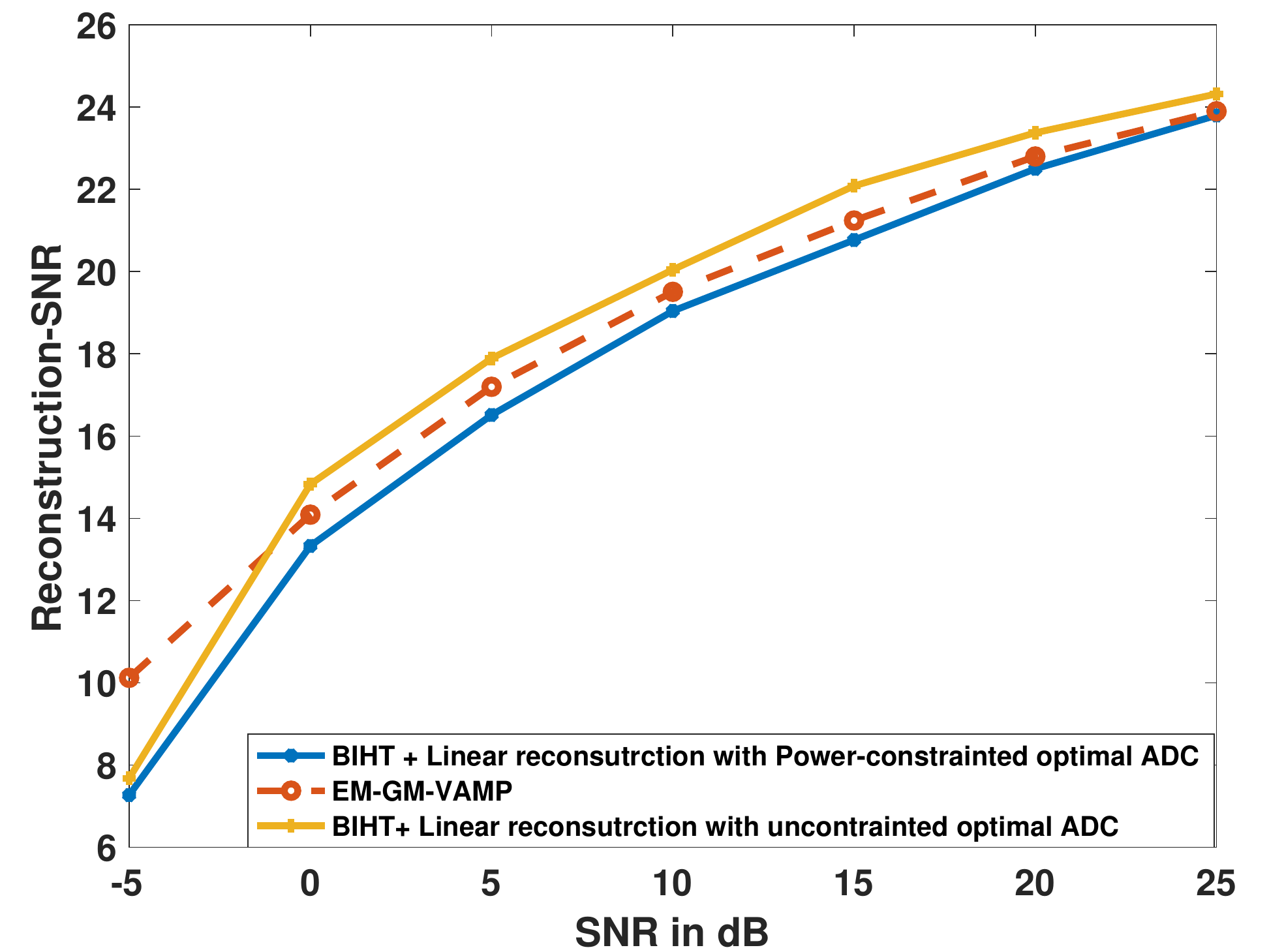}
\caption{$N_t=N_r=64$; training length of BIHT+ Linear reconstruction is 500, EM-GM-VAMP training length is 2048; delay spread $N=16$, and cluster number is set as 4.}
\label{fig:comp_mo}
\end{figure}

We provide a comparison between our BIHT +linear reconstruction method and results using EM-GM-VAMP from \cite{mo2018channel} for channel estimation as a function of ADC bit-depth. For direct comparison, we used the same channel model as in \cite{mo2018channel}; 
the multi-path channel has delay spread of $N=16$ and number of clusters equal to 4. As a result, the sub-channel vectors are {\em not independent} as in previous examples but correlated by clustering. 
As before our method has $20$ mW power budget and training length of 500 for each antenna, while EM-GM-VAMP imposes no power constraints and uses training length of 2048.
We observe in Fig.\ref{fig:comp_mo} that the BIHT plus linear reconstruction channel estimation performance is nearly the same as EM-GM-VAMP for the range of SNR shown, resulting in significant power savings. When the power budget is increased to $50$ mW, the ADC is able to run with higher sampling frequency and more bit-depth and our method (BIHT plus linear reconstruction) outperforms EM-GM-VAMP.

\begin{figure}[h]
  \centering
  \includegraphics[width=80mm, height=35mm]{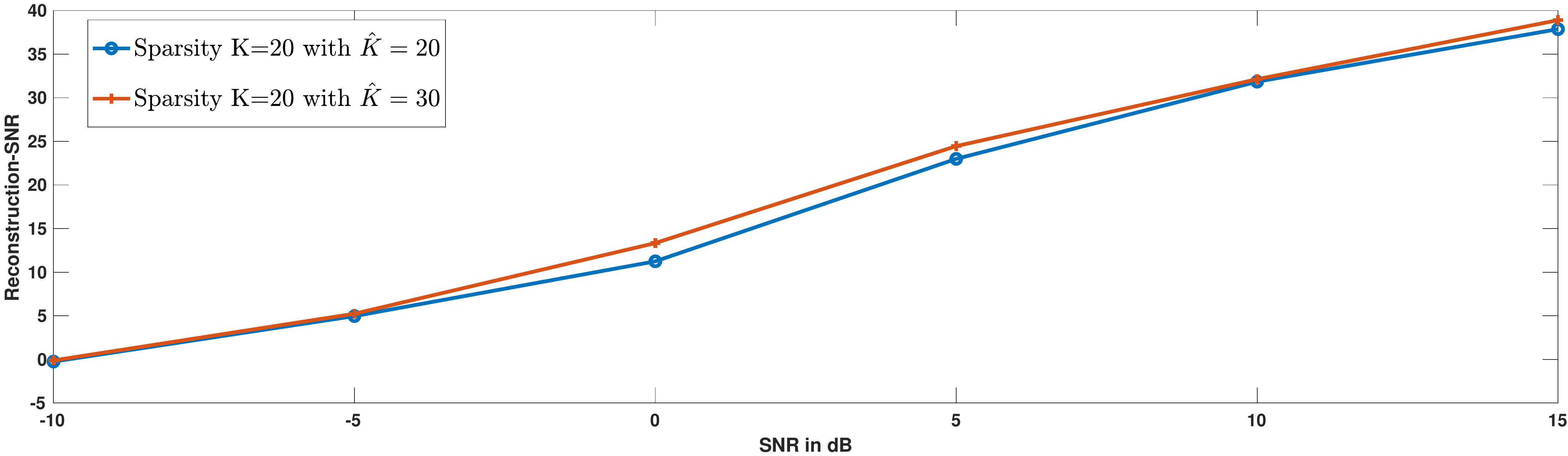}
\caption{$N_t=4$ and $N_r=2$. $\hat{K}$ is the heuristic value for BIHT and the channel estimation performance is robust if $\hat{K}$ is chosen to be larger than the real sparsity $K$.}
\label{fig:K}
\end{figure}

BIHT algorithm requires knowledge of sparsity $K$ since in Eq.\eqref{eq:BIHT} {\em the largest} $K$ terms are retained. 
 If in the algorithm a value $\hat{K}$ greater than the true value $K=20$ is chosen, the inherent robustness of compressed sensing kicks in, with no impact on resulting RSNR as shown in From Fig.\ref{fig:K}. 

\section{Conclusions \& Future Work}
In this work, a model for massive hybrid MIMO systems with few bit ADC was introduced that incorporates ADC power budget vis-a-vis the channel recovery algorithm that enables exploration of desirable operating points in terms of optimal parameters (ADC bit-depth and sampling rate). BIHT plus linear reconstruction method was identified as the desired approach and achievable error (RNSR) bounds based on appropriate training sequence design derived. Our results show that our proposed method significantly outperforms the traditional least-square approaches. Further, our results conclusively show that few-bit ADC (i.e. low power consumption) designs are feasible for our proposed CS-based channel estimation algorithm - i.e. the latter achieve very good channel RSNR, as compared to the optimal.  Since our approach only exploits channel time-domain sparsity properties and ignores any other prior knowledge - such as spatio-temporal correlations arising from clustered multi-path propagation - future work will seek to incorporate such features to further improve channel recovery with limited ADC precision.

\appendix{}
\subsection{Binary iterative hard thresholding (BIHT)} \label{BIHT}
Let $\mathbf{\tilde{y}}_1 =sgn(\mathcal{X} \mathbf{\tilde{h}}+\mathbf{\tilde{e}})$, where $\mathbf{\tilde{e}}$ has i.i.d. elements $\mathbf{\tilde{e}}_i \sim N(0,\sigma^2)$, $\mathbf{\tilde{h}} \in \mathbf{R}^{N}$ is the channel vector with sparsity of $K$ to be recovered, and $\mathbf{\tilde{y}}_1 \in \mathbf{\{-1,+1\}}^{N_rM}$ is the signal quantized to one bit. Note that all matrices and vectors are real. The BIHT algorithm consists, post Initialization: $\mathbf{\tilde{h}}^0=0$: of the following iteration: \\
\begin{equation}\label{eq:BIHT}
    \mathbf{a}^{l+1}=\mathbf{\tilde{h}}^l+\frac{\tau}{2}\mathcal{X}^T(\mathbf{\tilde{y}}_1-sgn(\mathcal{X} \mathbf{\tilde{h}}^l))
\end{equation}
where $\mathbf{\tilde{h}}^{l+1}=\eta_{\hat{K}}(\mathbf{a}^{l+1})$, and $\eta_{\hat{K}}(.)$ is a function which selects $\hat{K}$ largest elements in magnitude by thresholding. In general,  $\hat{K} \, \ge\, K$ suffices and Eq.~\ref{eq:BIHT} is repeated until convergence, and the resultant normalized onto the unit $l_2$ sphere. The step-size $\tau$ is a heuristic value for gradient descent step size which, according to \cite{jacques2013robust}, could be chosen as $\frac{1}{\sqrt{N_rM}}||\mathcal{X}||^2_2$. In our simulations, we set the value of $\tau = 10^{-5}$.

 \begin{figure}[h]
  \centering
  \includegraphics[width=100mm, height=22mm]{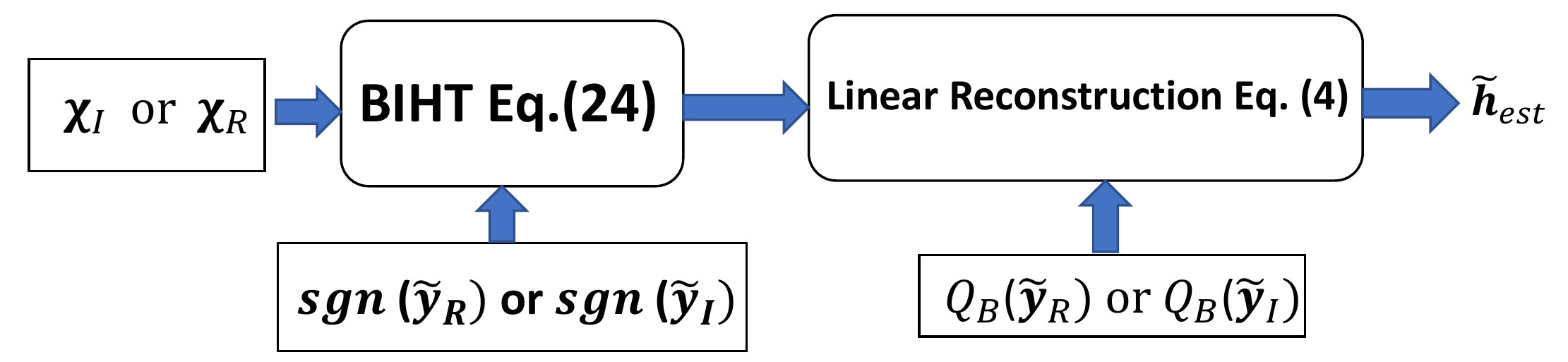}
\caption{Implementation diagram for BIHT plus Linear Reconstruction method: BIHT provides the non-zero support of channel vector to Linear reconstruction.}
\label{fig:csdiagram}
\end{figure}

\bibliographystyle{IEEEtran}
\bibliography{reference}
\end{document}